\documentclass[twocolumn, aps, prl, superscriptaddress]{revtex4-2}

\usepackage{mathptmx}
\usepackage{amsmath}
\usepackage{amssymb}
\usepackage{chemformula} 
\usepackage[table,x11names]{xcolor}
\usepackage{hyperref}

\usepackage{tikz}
\usetikzlibrary{shapes.geometric, shapes.misc}

\DeclareRobustCommand{\custommarker}[1]{%
  \begin{tikzpicture}[baseline=-0.6ex, line width=0.2mm,inner sep=0pt, outer sep=0pt]%
    #1
  \end{tikzpicture}%
}

\DeclareRobustCommand{\triup}[2]{\custommarker{\filldraw[draw=#1, fill=#2] (0,2.5pt) -- (-2.5pt,-1.5pt) -- (2.5pt,-1.5pt) -- cycle;}}
\DeclareRobustCommand{\tridown}[2]{\custommarker{\filldraw[draw=#1, fill=#2] (0,-2.5pt) -- (-2.5pt,1.5pt) -- (2.5pt,1.5pt) -- cycle;}}
\DeclareRobustCommand{\mysquare}[2]{\custommarker{\filldraw[draw=#1, fill=#2] (-2.1pt,-2.1pt) rectangle (2.1pt,2.1pt);}}
\DeclareRobustCommand{\mycircle}[2]{\custommarker{\filldraw[draw=#1, fill=#2] (0,0) circle (2.3pt);}}
\DeclareRobustCommand{\mystar}[2]{\custommarker{%
    \node[star, star points=5, star point ratio=2.25, draw=#1, fill=#2, inner sep=1.0pt, line width=0.2mm] at (0,0) {};}}

\begin{document}

\newcommand{\impmc}{Sorbonne Université, CNRS UMR 7590, Muséum National d'Histoire Naturelle, Institut de Minéralogie, de Physique des Matériaux et de Cosmochimie, IMPMC, 75005 Paris, France}
\newcommand{\lpens}{Laboratoire de Physique de L’\'{E}cole Normale Sup\'{e}rieure de Paris, CNRS, ENS \& Universit\'{e} PSL, Sorbonne Universit\'{e}, Universit\'{e} Paris-Cit\'{e}, 75005 Paris, France}

\title{Pressure–Temperature Phase Diagram and $\lambda$-Transition in Liquid Sulfur}

\author{Sonia Salomoni}
\email{sonia.salomoni@phys.ens.fr}
\affiliation{\lpens}
\affiliation{\impmc}
\author{Frédéric Datchi}
\affiliation{\impmc}
\author{A. Marco Saitta}
\email{marco.saitta@phys.ens.fr}
\affiliation{\lpens}
\author{Arthur France-Lanord}
\email{arthur.france-lanord@cnrs.fr}
\affiliation{\impmc}

\begin{abstract}

Using molecular dynamics simulations driven by a machine-learned interatomic potential, we investigate at low to intermediate pressures the $\lambda$-transition of sulfur, a temperature-induced polymerization. At ambient pressure, we capture the melting of crystalline cyclo-octasulfur into a liquid of molecular rings. 
Within this liquid, the concentration of non-\ch{S8} rings increases with temperature; we show that these molecules act as reactive centers, which eventually trigger polymerization. 
We reproduce key experimental signatures of  the $\lambda$-transition,
including the sharp increase in heat capacity and the pronounced dependence of the transition temperature on the heating rate. 
Building on this, we reconstruct a phase diagram of polymerization up to intermediate pressures. Our results reveal a moderate decrease of the polymerization temperature with pressure, culminating with its merging with the melting line at a critical point.
Beyond this point, we provide direct evidence of polymerization emerging from the crystalline phase. 
By analyzing temperature-ramp trajectories, we observe the formation of non-\ch{S8} rings, open chains, and extended polymeric structures which retain features of the crystalline arrangement; further heating the system leads to disorder taking over through melting. Polymerization is therefore initiated slightly before melting. 
Altogether, our findings provide a microscopic picture of the $\lambda$-transition throughout the sulfur phase diagram.

\end{abstract}

\maketitle

Elemental sulfur exhibits a rich phase diagram, characterized by a remarkable molecular diversity, spanning cyclic rings and linear polymeric chains~\cite{steudel2003elemental}.
At low pressure and temperature -- below 10~GPa and 600~K -- the stable crystals are motifs of the \ch{S8} molecular ring~\cite{rettig1987refinement,templeton1976crystal} (orthorhombic $\alpha$-\ch{S8} and monoclinic $\beta$-\ch{S8}), while at higher pressure, the favored solid structures are based on helicoidal polymeric chains~\cite{crichton2001situ,fujihisa2004spiral,degtyareva2005novel} (trigonal and tetragonal phases), or \ch{S6} rings with a very packed arrangement (rhombohedral sulfur)~\cite{donohue1961crystal,steidel1978redetermination,Crapanzano2005}.

A similar competition between ring-like and polymeric-like molecules has also been observed in the liquid phase. At ambient pressure, sulfur melts at $\text{T}_m$ = 388~K into a mobile, yellow liquid consisting primarily of \ch{S8} rings. Above $\text{T}_\lambda \approx$ 432~K, these rings are known to open (Eq.~\ref{eq:activation}) and polymerize (Eq.~\ref{eq:catenation}) to form long linear chains, fundamentally altering the liquid's structure and dynamics~\cite{gee1952molecular,tobolsky1959equilibrium,koningsberger1971polymerization}:
\begin{gather}
\label{eq:activation}
     \ch{S8 <=> ^{.}S8^{.}} \\
\label{eq:catenation}
    \ch{^{.}S_i^{.} + ^{.}S_j^{.} <=> ^{.}S_{i+j}^{.} }
\end{gather}

The propagation of polymerization is thermodynamically driven by entropy overcoming enthalpy in the competition between small cycles and polymeric chains.
This conceptual framework has been supported early on by experimental evidence from electronic spin resonance spectroscopy, differential thermal analysis, and static magnetic susceptibility measurements~\cite{poulis1963lengths,koningsberger1971polymerization,klement1974study}; it was confirmed later using vibrational spectroscopy~\cite{mausle1981molekulare}, high-performance liquid chromatography~\cite{steudel1985quantitative}, and X-ray diffraction~\cite{bellissent1990liquid,bellissent1994polymerization,crapanzano2006polymorphism}.
As pressure increases, $\text{T}_\lambda$ either slightly decreases~\cite{eisenberg1963equilibrium} or plateaus~\cite{kuballa1971differential}; it eventually intersects the melting curve.
This intersection was predicted by Eisenberg~\cite{eisenberg1963equilibrium} to occur near 0.086~GPa, and measured experimentally between 0.07 and 0.13~GPa~\cite{cova1953effect,vezzoli1969high,kuballa1971differential,brollos1974optical}.
Above this pressure, crystalline sulfur is expected to melt directly into a polymeric liquid.

A peculiar characteristic of polymerization in liquid sulfur is the remarkable sharpness of its onset, which has earned it the denomination of “$\lambda$-transition"~\cite{gee1952molecular}. Indeed, polymerization is marked by sudden, dramatic changes in physical properties such as color, viscosity~\cite{bacon1943viscosity,sofekun2018rheology} and heat capacity~\cite{west1959heat,feher1971chemistry}, while also inducing notable, but less pronounced, variations in density~\cite{zheng1992density} and diffusivity~\cite{saxton1953diffusion}.
Statistical physics treatments, including renormalization group approaches accounting for fluctuations and non-mean-field features, have been applied over the years~\cite{wheeler1980equilibrium,wheeler1983bicriticality,petschek1986equilibrium}, but could never reproduce experimental results satisfactorily~\cite{greer1998physical}.

Modeling this system from first principles is particularly demanding.
Molecular dynamics (MD) simulations based on density functional theory (DFT) are computationally prohibitive for polymeric liquids, which require large system sizes and long time scales: experimentally, equilibration near $\text{T}_\lambda$ takes approximately one hour~\cite{klement1970polymer}, and confinement studies suggest a minimum volume of 200~nm$^3$ to approximate (and 4000~nm$^3$ to recover) bulk behavior~\cite{begum2013modeling}. Despite these constraints, DFT calculations have been applied to study depolymerization mechanisms in the gas phase~\cite{kemper2020depolymerization}, and to characterize small (480-atom) model melts near the $\lambda$-transition~\cite{flores2022crucial}.

\begin{figure*}[ht!]
\centering
\includegraphics[width=\textwidth]{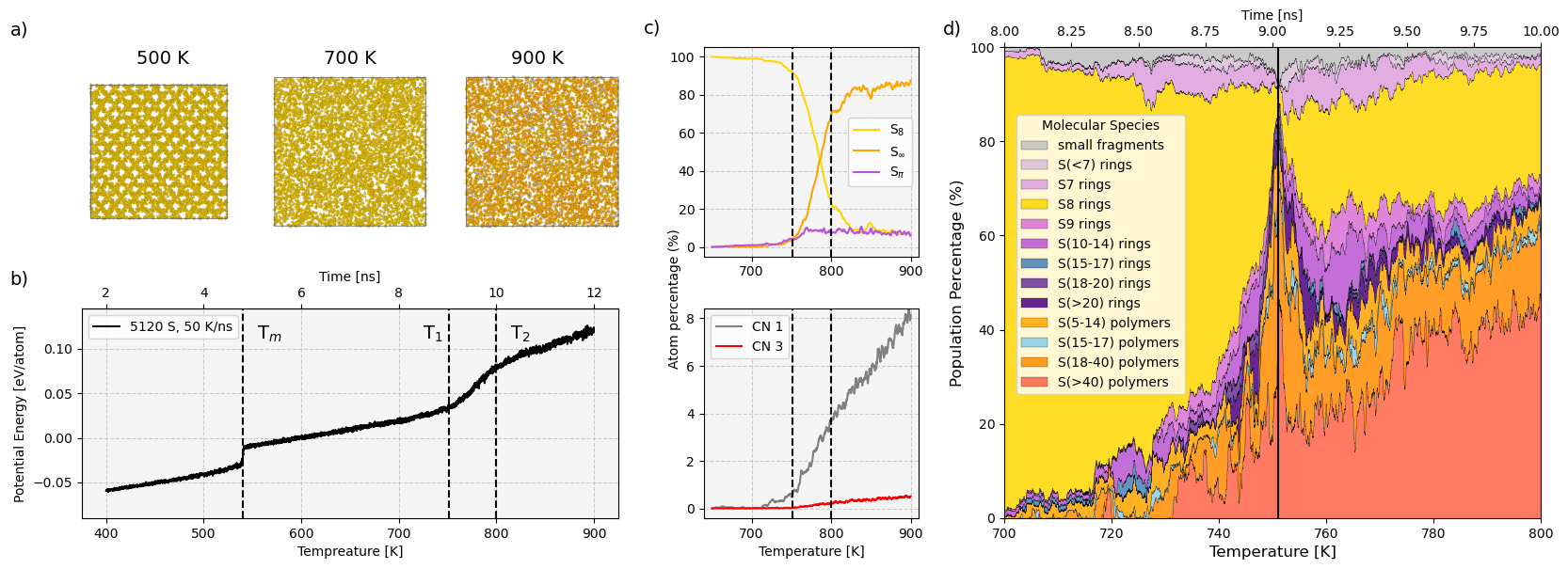}
\caption{
\textbf{Temperature ramp from solid $\alpha$-\ch{S8}, at 1~atm and 50~K/ns. 
}\textbf{(a)} Visual representations of the system at selected temperatures according to the following color coding: grey for 1-coordinated atoms, yellow for 2-coordinated atoms in ring molecules, orange for 2-coordinated atoms in polymers, red for 3-coordinated atoms).
\textbf{(b)} Time evolution of the atomic potential energy.
\textbf{(c)} Time evolution of the speciation, using a distance cutoff of 2.5~\AA. Upper panel: fraction of \ch{S8} rings, (\ch{S8}), of polymeric chains ($\text{S}_{\infty}$), and of non-\ch{S8} rings between sizes 5 and 20 ($\text{S}_{\pi}$); bottom panel: atomic coordination. 
\textbf{(d)} Stacked plot of the speciation lineage of polymeric atoms at the onset of polymerization. The solid line marks the reference temperature/time step: left of this line shows the provenance of polymeric atoms, while the right part represents their fate.   
}
\label{fig:1}
\end{figure*}

Empirical interatomic potentials can circumvent size and sampling limitations, but accurately capturing bond-breaking and -forming events remains challenging~\cite{muser2023interatomic,van2024methods}.
Ballone~\textit{et al.}~\cite{ballone2003density} pioneered Monte Carlo simulations of the liquid above $\text{T}_\lambda$ using a potential involving both bonded and non-bonded terms,  adjusted to DFT data~\cite{jones2003density}. More recently, Wang~\textit{et al.}~\cite{wang2025atomic} explored polymerization with a ReaxFF potential. Machine-learned interatomic potentials (MLIPs) offer a more systematic path: they learn potential energy surfaces from expressive basis functions without manual parameter tuning. Notably, Yang~\textit{et al.}~\cite{yang2024structure} combined a DeePMD model with enhanced sampling and learned collective variables to simulate reaction mechanisms during a transition between purely molecular and purely polymeric phases.

Several fundamental questions about sulfur's phase diagram remain open at the atomistic level.
In particular, the full sequence of transformations from crystalline $\alpha$-sulfur through molecular melting to the $\lambda$-transition has never been directly simulated.
Another challenge is reproducing the heating rate and pressure dependence of $\text{T}_\lambda$, as observed experimentally, alongside identifying the microscopic mechanisms driving partial polymerization at this transition; whether they align with the hypotheses of phenomenological models~\cite{tobolsky1959equilibrium}, with observations on transitions between ideal phases using MLIPs~\cite{yang2024structure}, or made with empirical potentials~\cite{wang2025atomic}.
Finally, it is of particular interest to understand how the transition proceeds in the high-pressure regime where the melting and polymerization lines coincide.

In this work, we address these questions using large-scale MD simulations driven by a local equivariant MLIP trained on a DFT dataset representative of both molecular and polymeric phases, with structures selected through active learning.
We explore the phase diagram \textit{via} temperature ramps (T-ramps) at fixed pressures, along which both the melting and the $\lambda$-transition spontaneously occur. We observe a strong dependence on heating rate of the lambda transition temperature, in line with experimental observations~\cite{klement1974study}. 
At ambient pressure, we identify the formation of non-\ch{S8} molecular rings as seeds that trigger polymerization, confirming a long-standing hypothesis~\cite{steudel2003liquid}.
At higher pressure, we observe that non-\ch{S8} cycles and polymeric chains already form in the pre-melting solid, conforming to the lattice, and that melting subsequently initiates from these polymeric portions -- providing the first atomistic picture of the coincident melting and polymerization regime.

We trained an Allegro model~\cite{musaelian2023learning} using a dataset at the spin-polarized DFT GGA-PBE~\cite{perdew1996generalized} level of theory. Configurations were generated via \textit{ab initio} MD at target temperature and pressure conditions, then selected using an active learning strategy. The model was validated against equations of state of crystalline phases outside the training set, and assessed for stability during extended liquid-phase simulations. 
Full details on data generation and selection, MLIP training and validation, and MD simulation settings are provided in the End Matter and Supplemental Material Sec.~A--F~\cite{suppmat}.

We first analyze a T-ramp initiated from an $\alpha$-\ch{S8} crystal (5120 atoms, or 640 rings) at 1~atm (Fig.~\ref{fig:1}). 
The system is heated from 300~K to 900~K in 12~ns, thus at a heating rate of 50~K/ns. Although the latter is many orders of magnitude faster than reported experimental conditions, it allows us to observe both melting and polymerization within a reasonable simulation time. Both a qualitative visual inspection [Fig.~\ref{fig:1}~\textbf{(a)}] and a quantitative topological analysis based on atomic coordination (using a 2.5~\AA\, cutoff distance) [Fig.~\ref{fig:1}~\textbf{(c)}] identify the energy discontinuity [Fig.~\ref{fig:1}~\textbf{(b)}] near $\text{T}_m$ = 540~K as melting into an \ch{S8} molecular liquid, and the change in slope between $\text{T}_1$ = 751~K and $\text{T}_2$ = 800~K as the onset and endset of polymerization ($\text{T}_1$ and $\text{T}_2$ were obtained \textit{via} a piece-wise linear fitting of the potential energy curve above $\text{T}_m$).

\begin{figure*}[t!]
\centering
\includegraphics[width=\textwidth]{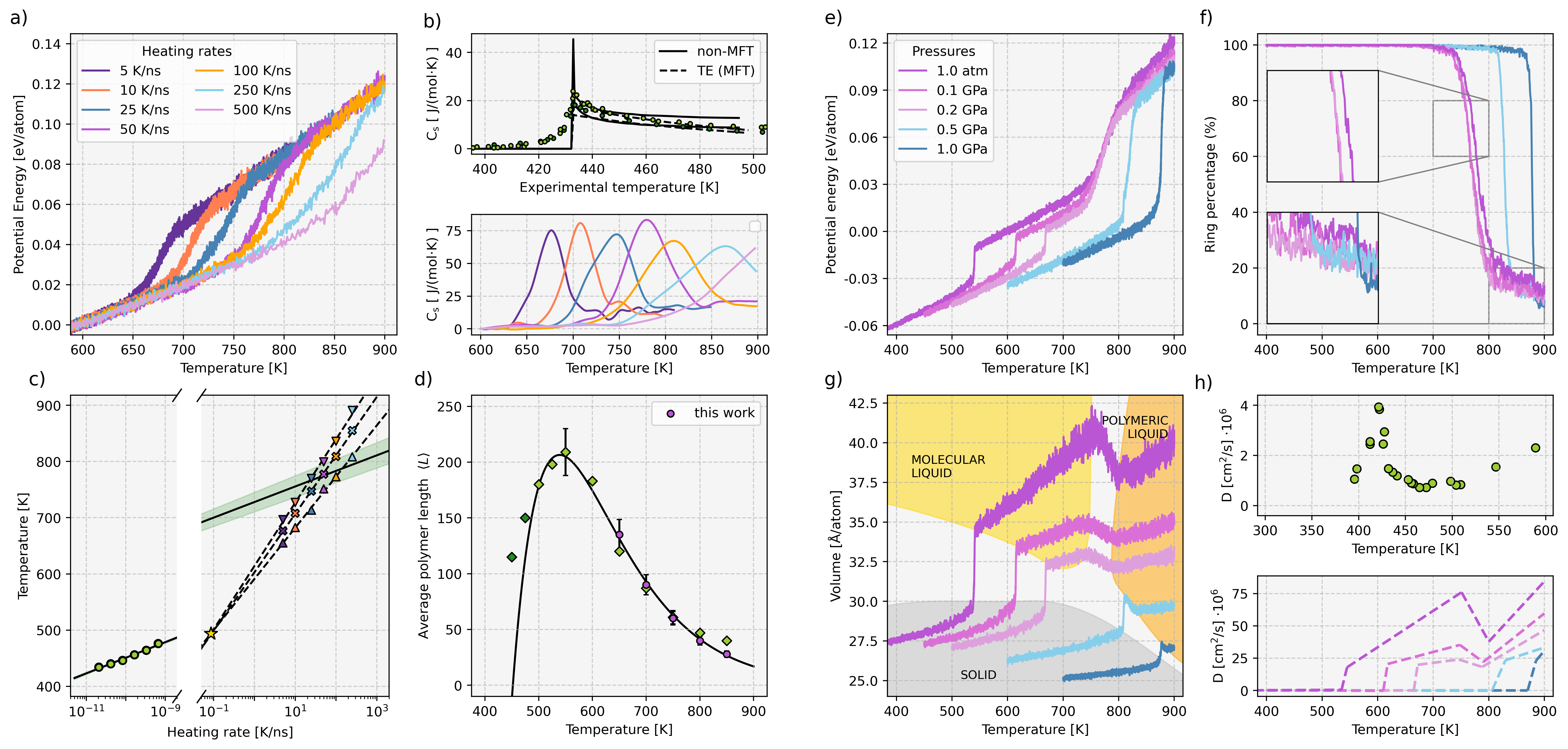}
\caption{
    \textbf{Heating rate and pressure dependence.}
    \textbf{(a)} Time evolution of the potential energy, for various heating rates. 
    \textbf{(b)} Singular part of the heat capacity $\text{C}_\text{s}$ as a function of temperature: in the upper panel, experimental measurements (\mycircle{black}{SeaGreen3}~\cite{feher1971chemistry}, \mycircle{black}{OliveDrab2}~\cite{west1959heat} ) and theoretical predictions from a mean-field (TE) theory (dashed lines~\cite{tobolsky1959equilibrium}) and non-mean-field (non-MFT) theory (solid lines~\cite{wheeler1980equilibrium}) with two different parametrizations~\cite{greer1998physical}; in the bottom panel, calculated from the T-ramps, with the same heating-rate color code as in panel (a).
    \textbf{(c)} Onset (\triup{white}{black}) and endset (\tridown{white}{black}) of the polymerization temperature as a function of the heating rate, with logarithmic fits (black dashed lines), and extrapolated $\text{T}_{\lambda}$ from the crossing of the fit lines (\mystar{black}{yellow}). Experimental data~\cite{klement1974study} (\mycircle{black}{OliveDrab3}) with their logarithmic fit (black solid line) are also reported. 
    \textbf{(e)} Potential energy, \textbf{(f)} volume, \textbf{(g)} ring content and \textbf{(h)} diffusivity (multi-segment fit from the raw data) evolution during simulations at a 50~K/ns heating rate and selected pressures. Symbols (\mycircle{black}{OliveDrab3}) in the diffusivity plot are experimental data~\cite{saxton1953diffusion}.
}
\label{fig:2}
\end{figure*}

As polymerization sets in, the fraction of atoms in ring molecules drops to roughly 20\% and continues declining slowly past $\text{T}_2$ to 5--10\%. Concurrently, the atomic fraction in open chains rises sharply between $\text{T}_1$ and $\text{T}_2$, reaching over 80\%.
The experimental polymer content, as estimated from Raman spectroscopy, saturates near 70\% in bulk system and at lower values under confinement~\cite{kalampounias2003rounding}. 
We also observe a significant population of rings with sizes 6--20 (denoted S$_\pi$), growing rapidly to 5--10\% around $\text{T}_1$ and remaining roughly constant thereafter. The fraction of 1-coordinated atoms, linked to radical chain ends, increases linearly with temperature, initially fueled by ring opening at $\text{T}_1$ then by depolymerization into shorter chains beyond $\text{T}_2$. The modest increase in 3-coordinated atoms originates from metastable tadpole molecules~\cite{stillinger1986chemical} and branched polymers. 

The S$_{\pi}$ content has been linked to the onset of polymerization~\cite{steudel2003liquid}: non-\ch{S8} rings exhibit lower dissociation energies than cyclo-octasulfur and would naturally serve as precursors to open chains that subsequently catenate. We verify this hypothesis by tracking the lineage of all polymeric atoms at a reference time step corresponding to $\text{T}_1$ [Fig.~\ref{fig:1} \textbf{(c)}]. While all polymeric atoms can ultimately be traced back to \ch{S8} rings, non-\ch{S8} species -- in particular \ch{S7} -- are strongly over-represented close to the onset of polymerization relative to their bulk population, while \ch{S16} are not. This analysis directly confirms that non-\ch{S8} species play a critical role in the polymerization mechanism, and challenges the original Tobolsky--Eisenberg picture of \ch{S8} chain-opening and catenation~\cite{tobolsky1959equilibrium}. Tracking the subsequent fate of polymeric atoms (positive times in Fig.~\ref{fig:1}~\textbf{(c)}) shows that many recombine into \ch{S8} rings, illustrating the reversible and complex character of the transformation. We do not observe the large macrocycles recently reported with ReaxFF~\cite{wang2025atomic}; differences between both models are discussed in Supplemental Material Sec.~J~\cite{suppmat}.

Next, we analyze T-ramps starting from the \ch{S8} molecular liquid at $\text{T} = 600$~K, with heating rates from 500 to 5~K/ns (simulation times: 0.6--60~ns).
The potential energy curves (Fig.~\ref{fig:2}~\textbf{(a)}) all exhibit the S-shape variation associated with the lambda transition, with a systematic shift to higher temperature with increasing heating rate. We also report the singular part of the heat capacity $C_\text{s}$, computed as the first derivative of potential energy with respect to temperature, smoothed with a Savitzky-Golay filter (25~K window, second-order polynomial), compared to experimental measurements~\cite{west1959heat,feher1971chemistry} (Fig.~\ref{fig:2}~\textbf{(b)}). Both datasets are plotted near their respective apparent transition regimes. We qualitatively reproduce the characteristic $\lambda$-shaped peak. In addition, the peak height is closer to the non-mean-field predictions of Wheeler~\cite{wheeler1980equilibrium} rather than to mean-field theory~\cite{tobolsky1959equilibrium}.
Precise experimental measurement of the heat capacity near the transition is challenging and strongly heating-rate dependent~\cite{ward1969investigation}; consistently, reducing the heating rate in our simulations would concentrate the peak of $\text{C}_\text{s}$ in a narrower temperature range, approaching a more critical behavior.

Then, we determined $\text{T}_1$ and $\text{T}_2$ in the same manner as above (see Supplemental Material Sec.~I~\cite{suppmat} for temperature values) and found that they approximately follow a logarithmic dependence on heating rate  [Fig.~\ref{fig:2}~\textbf{(c)}]. Extrapolating to lower heating rates down to the point where $\text{T}_1$ and $\text{T}_2$ coincide yields $\text{T}_\lambda \approx 500$~K, reasonably close to the experimental value of 432~K.
It is know from differential scanning calorimetry experiments~\cite{klement1974study} that $\text{T}_\lambda$ depends on the heating rate; the dependence appears to be weaker than in our simulations, which might be due to the very different rate regimes probed. 
We used the trajectory obtained employing the slowest heating rate (5~K/ns) to calculate the average polymer length (Fig ~\ref{fig:2} \textbf{(d)}). The value at 650~K was obtained following a 10~ns NPT equilibration to reach a converged polymer length distribution. Our results agree well with the Monte Carlo study of Ballone~\textit{et al.}~\cite{ballone2003density}, performed on a system of comparable size. Polymer length distributions are consistent with Flory theory~\cite{flory1936molecular} (Supplemental Material Sec.~G~\cite{suppmat}), and display finite-size dependence as already observed in confined experimental studies~\cite{kalampounias2003rounding, begum2013modeling}.

The last set of simulations comprises T-ramps starting from $\alpha$-\ch{S8} at pressures up to 1.0~GPa, at a fixed heating rate of 50~K/ns. Time evolutions of potential energy, specific volume, ring content, and diffusivity are reported in Fig.~\ref{fig:2} \textbf{(e)--(h)}. Up to 0.2~GPa, melting into a molecular liquid is clearly identified by an abrupt increase in potential energy, specific volume, and diffusivity at constant 100\% ring content. Polymerization is instead characterized by a slower energy increase, a volume decrease of 3--5\%, and a significant drop in diffusivity -- effects that are most pronounced at ambient pressure. 
We observe a small $T_\lambda$ downshift of approximately 10~K at 0.1 and 0.2~GPa relative to ambient pressure, consistent with a weakly negative slope of the polymerization line. Experiments either suggest a similar slope~\cite{kuballa1971differential}, or almost no slope at all~\cite{brollos1974optical}. We also report at 900~K a small but non-negligible decrease in the ring content at 1.0~GPa, which drops at about 7\%, with respect to lower pressures (10--11\%), which might be a hint of the liquid-liquid transition proposed by Henry \textit{et al.}~\cite{henry2020liquid}.
Regarding the diffusivity evolution, the shape correctly reproduce the decreased mobility associated with the formation of polymer chains, however the intensity difference with experimental observation~\cite{saxton1953diffusion} is likely due to out-of-equilibrium estimation.
At higher pressures (0.5 and 1.0~GPa), melting and polymerization are coupled: a single jump appears in the energy and diffusivity, intermediate in speed between the sharp melting at low pressure and the slower polymerization. In the volume evolution, the rapid expansion associated with melting is immediately followed by a gradual contraction as polymerization propagates.

\begin{figure}[t!]
\centering
\includegraphics[width=0.49\textwidth]{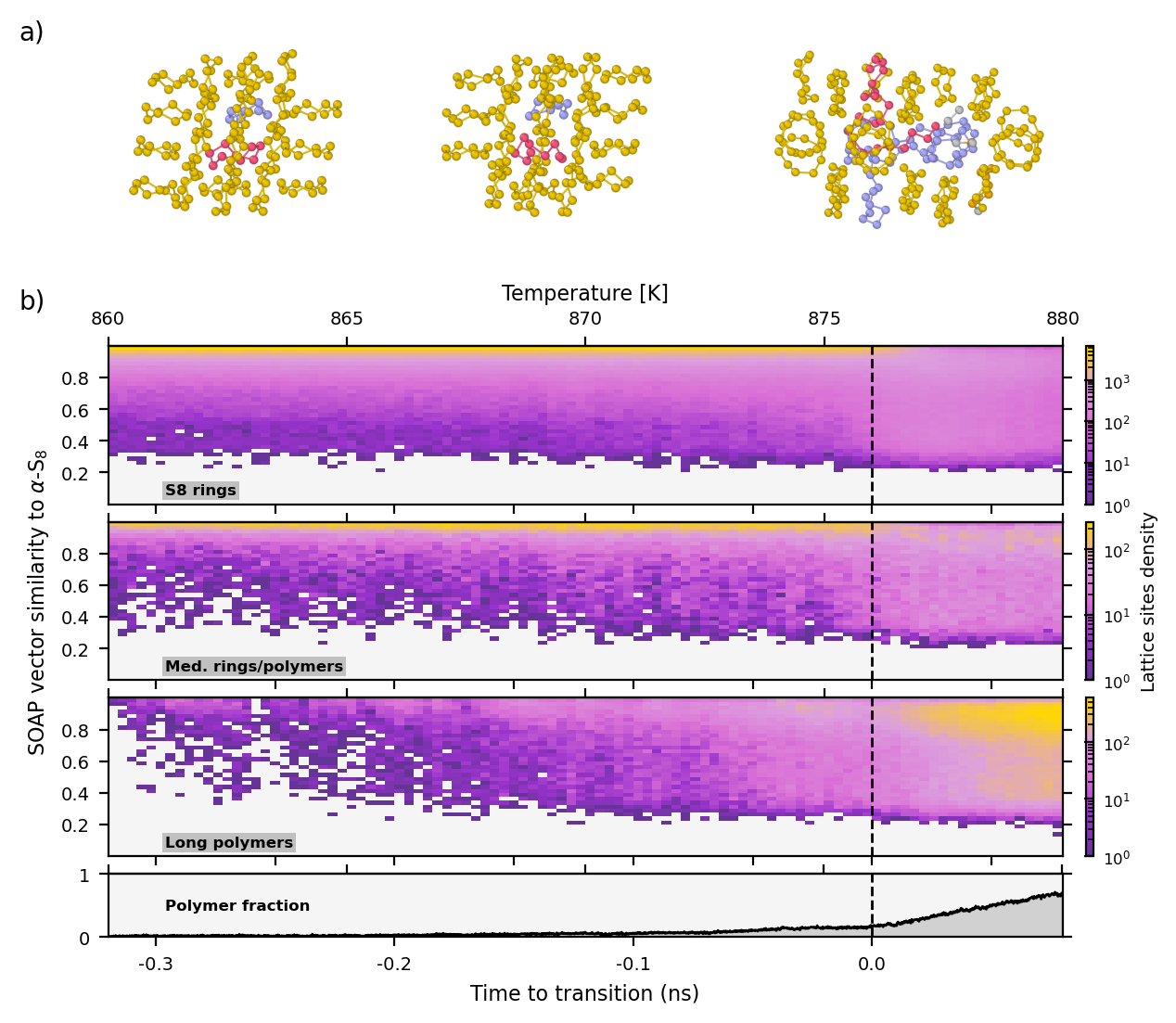}
\caption{
    \textbf{Analysis of the trajectory at 1.0~GPa with 50~K/ns heating rate.}
    \textbf{(a)} Visual representation of small rings, small polymers and long polymers conforming to the lattice in the pre-melting regime. The non-\ch{S8} fragments are highlighted with pink and purple colors.
    \textbf{(b)} SOAP vector similarity for \ch{S8}, small and long polymers in a selected temperature range across melting. 
    The vertical line marks the sharp increase in polymeric content and the concurrent broadening of the SOAP similarity distribution.
}
\label{fig:3}
\end{figure}

It is therefore interesting to further analyze the mechanism leading to melting and polymerization at these higher pressures. In particular, one would like to understand whether polymerization immediately follows melting, or precedes it in some degree. To do so, we performed a quantitative analysis over the 1.0~GPa T-ramp of the conformation of \ch{S8} rings, non-\ch{S8} rings and chains, and longer chains (formed by more than 13 atoms) within the crystalline lattice.
We identified molecular lattice sites as the centers of mass of \ch{S8} rings in the ideal orthorhombic structure, and computed a descriptor of the local structure in the form of a SOAP vector (Smooth Overlap of Atomic Position)  at each site~\cite{bartok2013representing}. The cosine similarity between each instantaneous SOAP vector and its perfect-crystal reference is plotted as time-dependent distributions in Fig.~\ref{fig:3}~\textbf{(b)}. For \ch{S8} rings, the distribution is sharply peaked near 1, with some thermal broadening; at the transition, it loses its unimodal character as lattice periodicity is destroyed by melting. Strikingly, both non-\ch{S8} rings and polymeric chains are present \textit{before} the transition, and their SOAP similarity distributions are peaked near 1, with progressive broadening before reaching the melting transition. This is particularly true for species occupying a single lattice sites (such as relatively small non-\ch{S8} rings), but also holds for longer polymeric chains. In Fig.~\ref{fig:3}~\textbf{(a)}, we show examples of small non-\ch{S8} rings, as well as small and long polymers conforming to the crystalline arrangement by coiling accordingly to the \ch{S8} rings within the lattice. We further discuss the occurrence of non-\ch{S8} rings and polymers conforming to the lattice in the pre-melting regime in Supplemental Material Sec.~K~\cite{suppmat}.
Polymerization is therefore initiated slightly before the melting, and accelerated at the transition. This form of polymerization within the crystal is to some extent akin to topochemical polymerization~\cite{hema2021topochemical}, in which polymerization is achieved in the solid state from an excitation, with no significant change in structure. 

\begin{figure}[t!]
\centering
\includegraphics[width=0.48\textwidth]{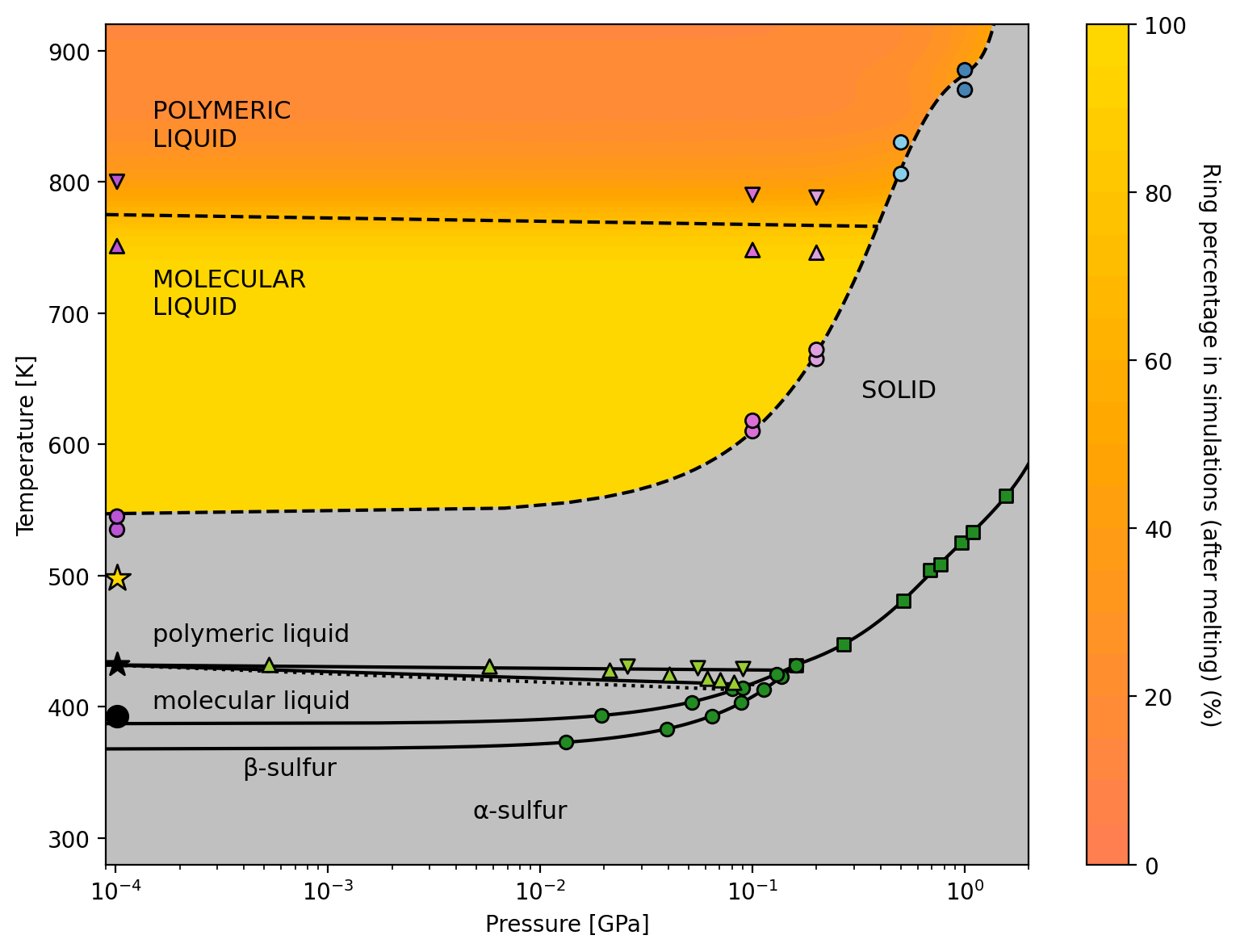}
\caption{
    \textbf{Experimental and computational phase diagrams.}
    Experimental data points are taken from references: 
    % melting and solid transition
    \mycircle{black}{SpringGreen4} and \mycircle{black}{black}~\cite{tammann1903kristallisieren} (see~\cite{eisenberg1963equilibrium}), 
    \mysquare{black}{SpringGreen4}~\cite{crapanzano2006polymorphism}, 
    % polymerization
    \triup{black}{OliveDrab3}~\cite{kuballa1971differential}, 
    \tridown{black}{OliveDrab3}~\cite{brollos1974optical}, 
    \mystar{black}{black}~\cite{steudel2003elemental}. The black solid lines are interpolations of the experimental data, and the dotted line is the theoretical prediction from~\cite{eisenberg1963equilibrium}. The symbol \mystar{black}{Gold2} reports the extrapolated computational polymerization temperature. The other symbols mark the onset and endset of polymerization and melting obtained during T-ramp simulations at 50~K/ns. Dashed lines are obtained from the linear fit or spline interpolation of these points.
}
\label{fig:PT}
\end{figure}

Finally, we recap our results into a pressure-temperature phase diagram (Fig.~\ref{fig:PT}), comprising both the literature experimental data and our computational results from the 50~K/ns T-ramp. The first obvious takeaway is that the strong dependency on heating rate of the $\lambda$-transition leads to a large shift in temperature between experiment and computational results. This can be alleviated to some extent: the extrapolated $\lambda$-transition temperature is also reported. Features of the phase diagram are reproduced, including the existence of a molecular liquid phase, the transition to a mixed-polymeric liquid, the melting temperature increase with increasing pressure, the weak dependence of the polymerization line on pressure, and the eventual merging of both transition phenomena. 
We mention that we did not attempt to capture is the solid-solid transformation from $\alpha$-\ch{S8} to $\beta$-\ch{S8} upon heating, as it's kinetically hindered-\cite{steudel2003solid}. As a complement to this phase diagram, the analysis of transition mechanisms enabled by atomic-scale simulations allows us to learn valuable insights regarding polymerization in sulfur. In particular, it appears that non-\ch{S8} rings act as epicenters of polymerization, and that at high-enough pressure, polymerization is initiated within the crystal lattice. There is room for improvement regarding the treatment of spin degrees of freedom~\cite{cuarare2024random}: even if we account for spin polarization, quantum chemistry calculations at a fixed multiplicity prevent us from properly describing very small fragments such as isolated sulfur atoms, which can show up in MD simulations driven by an MLIP. Appropriately capturing non-singlet states in the labeling and training phases, however challenging, would allow one to avoid having to preemptively forbid the formation of single-atom fragments. Finally, future research should be directed at modeling the liquid-liquid transition recently reported under moderate pressures~\cite{henry2020liquid}. 

We hope this study will motivate further experimental attempts to characterize the polymerization transition in order to compare with our predictions. In particular, \textit{in situ} Raman spectroscopy could further help understanding how polymerization proceeds from the crystal  at intermediate pressures. 

\textit{Acknowledgments ---}
This work has received support from the Sorbonne Center for Artificial Intelligence (SCAI). This work was granted access by GENCI to the HPC resources of CINES, IDRIS, and TGCC under the allocations A0160901387, A017081418, A0180901387, and A0190814184. The authors thank Mattia Perrone and Matteo Cioni for fruitful discussions regarding the crystalline similarity analysis, and Mohamed Mezouar and Laura Henry for discussions and suggestions.  

\textit{Data availability ---}
The trained model, input scripts, and raw data will be uploaded upon publication of this manuscript.

\bibliography{biblio}

\onecolumngrid
\vskip30pt

\begin{center}
\textbf{\large End Matter}
\end{center}
\label{endmatter}

\bigskip

\twocolumngrid

\textit{Appendix: AIMD simulations.} 
All AIMD simulations and single-point energies and forces calculations needed to train and test the machine learning interatomic potentials (MLIP) were performed using the  CP2K~9.2 code~\cite{CP2K2020}, and the Quickstep package~\cite{VandeVondele2005quickstep}.

To generate our 18 AIMD trajectories we chose the NVT ensemble and a time step of 2.0~fs. The initial configurations are generated by randomly filling a simulation box with 64~\ch{S8} molecules (corresponding to 480~S atoms), with three target densities~\cite{PACKMOL2019}:
$\rho_{1}$ = 1.84~g/cm$^3$, $\rho_{2}$ = 1.93~g/cm$^3$, $\rho_{3}$ = 2.01~g/cm$^3$, which correspond to cubic simulation boxes with side lengths: $l_1$ = 24.03~\AA, $l_2$ = 23.67~\AA, $l_3$ = 23.34~\AA.
Temperature was fixed at 650~K, 850~K and 900~K, and controlled using a chain of Nosé-Hoover thermostats~\cite{evans1985nose} with coupling constants of 200~fs. Each MD simulation is run for 200~ps.

We performed spin-polarized calculations~\cite{pople1995spin} (unrestricted Kohn-Sham) at a fixed multiplicity of 1 (singlet state); the exchange-correlation functional was approximated using the Generalized Gradient Approximation (GGA), as parametrized by Perdew–Burke–Ernzerhof (PBE)~\cite{perdew1996generalized}. The Kohn-Sham orbitals were expanded using a double-zeta valence (DZVP) Gaussian short-range (SR) molecularly optimized (MOLOPT) basis set~\cite{vandevondele2007gaussian}. The auxiliary plane wave expansion, used to efficiently compute the Hartree and exchange-correlation potentials with the Gaussian and Plane Waves method~\cite{VandeVondele2005quickstep} (GPW), was truncated at an energy cutoff of 600~Ry, with a relative energy cutoff of 60~Ry. These basis set choices are discussed in Supplemental Material  Sec.~A~\cite{suppmat}. The core electrons were treated using the Goedecker-Teter-Hutter (GTH) pseudopotential~\cite{goedecker1996separable}, optimized for PBE. We also included empirical dispersion interaction corrections, in the form of the D3 method~\cite{grimme2010consistent} to capture the long-range electron correlations. We sampled the Brillouin zone at the $\Gamma$-point only, since we used large simulation boxes. The spin-unrestricted Kohn-Sham equations were solved self-consistently using the orbital transformation method~\cite{VandeVondele2003efficient}. 

\textit{Appendix: data selection.} 
We first sampled the AIMD trajectories at 25-step intervals to reduce structural correlation. 
From this set, we employed a "query by committee" (QBC) active learning strategy~\cite{schran2020committee} to assemble a dataset. This utilized a committee of eight high-dimensional neural network potentials~\cite{behler2007generalized} (HDNNPs) trained using the n2p2 software~\cite{singraber2019library}.

Specifically, each atomic neural network features two hidden layers of 20 neurons each, utilizing $\tanh$ activation functions for internal nodes and a linear function for the output. To represent the atomic environments, we generated atom-centered symmetry functions using an automated procedure~\cite{behler2011acsf}. To ensure comprehensive radial and angular coverage, we employed 16 radial descriptors (shifted Gaussians) and 8 angular descriptors (concentric Gaussians), both utilizing a cutoff distance of 6.0~\AA. Network weights and biases were optimized via a parallel multi-stream Kalman filter~\cite{blank1994adaptive}. 

After an initial training on 100 randomly sampled configurations, we performed active learning iterations. In each cycle, up to 50 new structures were selected from a pool of 5000 candidates, based on committee disagreement (the standard deviation of the distribution of predictions considering the ensemble of models). The selection criteria combined force prediction uncertainty at 75\% and energy uncertainty at 25\%.
After 20 iterations, the QBC procedure allowed us to select a dataset comprising 970 structures. A topological comparison confirmed that this active learning approach captured a broader variety of configurations than an equivalent random sampling (see Supplemental Material Sec.~B~\cite{suppmat}).

\textit{Appendix: MLIP training.} 
Before initiating the training procedure, we subtracted the D3 dispersion correction contributions to the energies and forces of the QBC dataset.
We benchmarked different Allegro~\cite{musaelian2023learning} models. Since Allegro is a strictly-local equivariant message-passing architecture, the important hyperparameters to evaluate are the number of interaction layers ($T$), the number of equivariant channel ($L$), and the radial cutoff ($r_{cut}$). Our best model featured $T=4$ interaction layers, $L=2$ equivariant channels, and a $r_{cut}$ = 6.0~\AA\, cutoff. The hyperparameter selection process is reported in Supplemental Material Sec.~C~\cite{suppmat}. 

The settings related to the model architecture were the following: the radial chemical embedding multilayer perceptron (MLP), which generates the initial scalar latent features, consisted of 4 hidden layers of width 1024, connected by SiLu non-linearities and an output layer of dimension 128. The latent Allegro MLP consisted of 3 hidden layers of width 512 (SiLU non-linearities), using as input 128 scalar features and 64 tensorial features. The final readout energy MLP had one hidden layer of width 128. All MLP weights were initialized according to a uniform distribution.

The models were trained using a joint loss function based on energies and force errors: the per-atom MSE loss is calculated using a weighting of 1:1 for the energy and force term. The Adam optimizer was used for training with an on-plateau learning rate schedule: the initial value was set to 0.001, reduced with a patience of $20$ epochs and a decay factor of 0.6. The QBC dataset was split into a training set (80\%), a validation set (10\%), and a test set (10\%). The best model was selected as the one minimizing the validation loss. The comparison in energy and force components between the DFT reference data from the test set and the model eventually employed in our MD simulations is reported in Supplemental Material Sec.~D~\cite{suppmat}.
The model predictions were also compared with DFT calculations of the equation of state~\cite{murnaghan1944compressibility} for the orthorhombic and trigonal solid phases, as well as the transition pressure between them (see Supplemental Material Sec.~E~\cite{suppmat}). This is a challenging test, since there is no information in the training datasets from any crystalline phase. 

\textit{Appendix: MLIP-MD simulations.} 
We used the LAMMPS software~\cite{LAMMPS2022} to perform molecular dynamics using the best-performing model. 
To run the MLIP-MD simulations, we chose the same timestep as in AIMD (2.0~fs). Temperature and pressure targets employed a Nose-Hoover chain of thermostat and barostat, with 200~fs and 2000~fs damping constants, respectively. The thermostat is used either to fix the temperature of the system, or to increase it at a constant rate. 

We added D3 dispersion corrections analytically to the MLIP baseline, using the dispersion/d3 pair potential. In particular, we selected the Becke-Johnson (BJ) damping variant, and a global cutoff radius of 12~\AA.

We prevent the occurrence of non-physical isolated sulfur atoms using a constraint on the smallest coordination number in the simulation. We observed their formation in unconstrained simulations at high temperatures: properly describing them would require performing DFT calculations in non-singlet states, which would significantly complicate the training process. We therefore relied on the PLUMED plugin~\cite{tribello2014plumed} to constrain the minimum coordination number. The minimum coordination number was calculated smoothly using a modified softmax function. The applied constraint took the form of a steep harmonic lower-wall on the minimum coordination number, located at 0.9.

Complementary analyses performed on the MD trajectories are reported in Supplemental Material Sec.~F--K~\cite{suppmat}.

\newpage

\end{document}